\documentclass[preprint,12pt]{elsarticle}

\usepackage{graphicx}
\usepackage{amssymb}
\usepackage{amsmath}

\usepackage{lineno}

\journal{Journal Name}

\begin{document}

\makeatletter
\def\ps@pprintTitle{%
  \let\@oddhead\@empty
  \let\@evenhead\@empty
  \let\@oddfoot\@empty
  \let\@evenfoot\@oddfoot
}
\makeatother

\begin{frontmatter}


\title{Self-assembly of cylinder forming diblock copolymers on modulated substrates: a simulation study}



\author{Karim Gadelrab and Alfredo Alexander-Katz*\let\thefootnote\relax\footnote{*aalexand@mit.edu} }

\address{Department of Materials Science and Engineering. Massachusetts Institute of Technology, Cambridge, Massachusetts, 02139, United States.}

\begin{abstract}
Self-consistent field theory (SCFT) and strong segregation theory (SST) are used to explore the parameter space governing the self-assembly of cylinder forming block copolymers (BCPs) on a modulated substrate. The stability of in-plane cylinders aligning parallel or perpendicular to substrate corrugation is investigated for different barrier height and spacing for a weakly preferential substrate. Within the conditions of our simulations, the results indicate that cylinder alignment orthogonal to substrate undulation is promoted at low barrier height when substrate is preferential to minority block, independent of barrier spacing. Commensurability is shown to play a limited role in the assembly of orthogonal meshes. Parallel alignment is readily achieved at larger barrier height, near condition of commensuration between barrier spacing and polymer equilibrium period. This is particularly true when substrate is attractive to majority block. The interplay between barrier shape and substrate affinity can be utilized in nanotechnology application such as mesh creation, density multiplication, and 3D BCP morphologies.
\end{abstract}




\end{frontmatter}


\section{Introduction}
\label{S:1}

Block copolymers (BCPs) have been intensively studied for their ability to self-assemble into a plethora of nanoscale morphologies\textsuperscript{1}. Even at its simplest form, a linear diblock copolymer can produce different periodic structures namely, lamellar, cylindrical, spherical, and gyroid; depending on the proportions and the degree of incompatibility between its constituents\textsuperscript{2}. The native order of these periodic patterns are inherently limited in spatial extent. In order to improve long-range order and introduce engineered features at will, various strategies of directed self-assembly DSA have been proposed such as post and trench templating\textsuperscript{3-7}, chemical patterning of substrates\textsuperscript{8-10}, shear alignment\textsuperscript{11-13}, laser zone annealing\textsuperscript{14}, to mention a few. Furthermore, efforts to study BCP self-assembly on a non-flat substrate proved to be a viable approach to achieve order\textsuperscript{15-21}. Modulated substrates resulted in improved long-range order and control over domain orientation\textsuperscript{15-19}. Thus, understanding the parameters governing self-assembly of a BCP on a non-flat substrate has a direct application in extending polymer morphologies beyond the 2D patterns into 3D through multi-stacking\textsuperscript{20, 22, 23}. 

In earlier work\textsuperscript{23}, multi-stacking of BCP domains was investigated by the authors and coworkers. Elaborate mesh patterns emerged when a cylindrical forming BCP layer self-assembled on top of a previously etched, untreated cylindrical morphology of a larger feature size. The emergent pattern was stable regardless of the overall order of the bottom layer. Self-consistent field theory (SCFT) results showed that barrier height and chemical affinity of substrate were critical process parameters. Recent experimental work on a similar system demonstrated the ability of the top layer to align parallel to the bottom one\textsuperscript{24}. It was argued that orientation manipulation is achieved through control over film thickness. While no exact measurements were conducted, change in brightness in the final scanning electron microscopy (SEM) micrographs were employed to qualitatively indicate variations in film thickness. The accompanied analysis using SCFT concluded that incommensurability between the two layers dictated orientation preference. There, orientation orthogonal to the bottom corrugated layer was stabilized when the hexagonal packing of parallel orientation could not be accommodated within the film. Despite the large computational cell and parameter space explored, the demonstrated morphologies had vague resemblance to the accompanied experiments in both parallel and orthogonal morphologies. We believe that improper boundary conditions applied in SCFT producing effectively tri-layer system (fixed modulated substrate and top double layered BCP) derived conclusions that are of little relevance to the accompanied experimental work. 

In this work we expand on the parameter space of this system to unequivocally show what important parameters (or parameter combinations) are which dictate the orientation of the multilayered cylindrical system. We show regions of stability of both perpendicular and parallel structures for extended barriers as a function of substrate affinity, barrier height, and barrier spacing. We employ 3D SCFT calculations to construct phase diagrams of parallel and perpendicular orientations as a function of the explored parameter space. We corroborate SCFT results with strong segregation theory (SST) analysis.  Indeed, weak substrate affinity to minority block and shallow substrate modulation promote perpendicular orientation of top layer with limited effect of barrier spacing, explicitly showing that commensurability is not the driving parameter for the assembly of orthogonal meshes, as was previously indicated\textsuperscript{23}. Concurrent simulation work has also found similar conclusions to our work\textsuperscript{25}. Large barrier height with respect to a monolayer forming thin film promoted parallel orientation in the vicinity of commensuration. This is particularly true when substrate is attractive to majority polymer. The results are relevant to film thicknesses forming a single monolayer of BCP domains. Our findings are discussed given the body of experimental work available in literature.




\section{Simulating BCPs under confinement}

BCP domain orientation in 1D confinement is a problem that has been extensively studied in the context of lamellar domains in thin films and trench confinement. Striped domains (2D projection of lamellae or in-plane cylinders) can align either parallel or perpendicular to confinement surfaces. SST predicted degeneracy points in free energy exist at commensurate wall spacing and neutral wetting conditions, meaning that both perpendicular and parallel orientations are equally attainable\textsuperscript{26}. On the other hand, SCFT analysis showed that for intermediate degrees of segregation, only domains normal to confining surfaces are stable for commensurate neutral walls. A nonzero wall attraction is necessary to cause BCP domains to align parallel to confining surfaces even at commensurate wall spacing\textsuperscript{27, 28}. This behavior has been demonstrated at the weak segregation limit as well\textsuperscript{29}. For neutral walls, both blocks can exist in the vicinity of the confining surface. Chains aligning parallel to walls is an orientation that is compatible with chain’s conformation at the \textit{AB} interface for perpendicular domains. In addition, there is a reduction in interfacial tension near wall due to lower polymer density. These effects are believed to stabilize the perpendicular orientation over the parallel one.  
To demonstrate this behavior, a 2D SCFT simulation of lamellar structure (\textit{f} = 0.5 and $\chi$N = 17) is performed. Here, \textit{f} is the volume faction of block \textit{A}, $\chi$ is the Flory-Huggins parameter, and N is the degree of polymerization. Simulation details follow that in \textsuperscript{23}. Wall spacing D[R\textsubscript{g}] where R\textsubscript{g} is the chain radius of gyration, is continuously varied and striped structure that is either parallel or perpendicular to confining surfaces is seeded to simulation. In figure 1, Parallel orientation shows a semi-parabolic free energy $\Delta$\textit{F}[ \textit{nk\textsubscript{b}T}] that has a minimum at wall separation D[R\textsubscript{g}] commensurate with domain equilibrium spacing L\textsubscript{0}. Deviation from integer multiples of L\textsubscript{0} applies stretching energy on BCP chains raising the energy of the system. Perpendicular orientation shows a continuous decay in $\Delta$\textit{F}[ \textit{nk\textsubscript{b}T}] with increasing wall spacing. Note that perpendicular orientation is stable (lower $\Delta$\textit{F}[ \textit{nk\textsubscript{b}T}]) independent of D[R\textsubscript{g}] even at walls that are weakly affine to block \textit{A} (\textit{w\textsubscript{-}} = 2). Increasing walls affinity for \textit{A} (\textit{w\textsubscript{-}} = 4) reduces $\Delta$\textit{F}[ \textit{nk\textsubscript{b}T}] of the parallel morphology showing regions of stability of parallel orientation in the vicinity of commensurate wall spacing. These calculations are example of using interfacial energy to control domain orientation in simple 1D full confinement of striped structures. We will refer to surface fields that are unable to stabilize parallel orientation of striped morphology at the commensurate condition in 1D confinement as weakly attractive, and surface fields that can stabilize parallel orientation as strongly attractive. Such fields are employed in studying monolayers of cylinder forming BCP thin films.   

\begin{figure}[h]
\centering\includegraphics[width=1.0\linewidth]{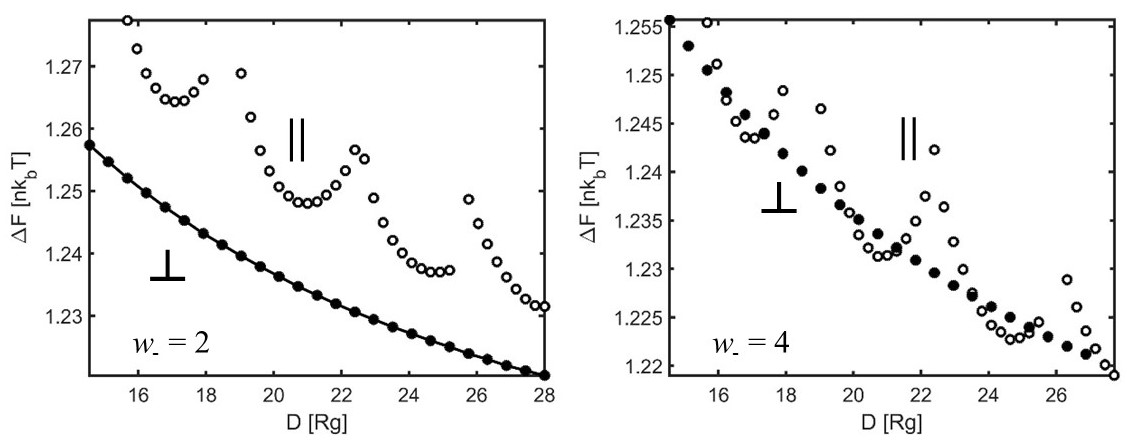}
\caption{Free energy as a function of wall spacing D[R\textsubscript{g}] for a lamellar morphology in 1D confinement. Open markers are for domains parallel to walls, while solid markers are for perpendicular orientation. At a non-zero wall attraction (\textit{w\textsubscript{-}} = 2), perpendicular orientation is stable even at commensurate wall spacing (left). Increasing wall attraction (\textit{w\textsubscript{-}} = 4) stabilizes parallel orientation when wall spacing is an integer multiple of polymer equilibrium periodicity (right).}
\end{figure}

SCFT simulations are conducted for an \textit{AB} diblock with a volume fraction \textit{f} = 0.33 and $\chi$N = 17. These simulations parameters result in the formation of cylindrical domains in 3D. Details of implementing SCFT are found elsewhere\textsuperscript{7, 23}. The unconstrained bulk simulations showed an equilibrium domain spacing L\textsubscript{B} = 3.96R\textsubscript{g}. Equilibrium domain spacing in thin film condition L\textsubscript{0} is expected to slightly differ from the bulk value due to the new symmetry imposed on the BCP domain\textsuperscript{30, 31}. In a monolayer cylindrical BCP thin film, in-plane cylinders adopt square symmetry parallel to the substrate. This is evident from the free energy plot of 2D simulation of circular domains confined between top surface and bottom substrate (see Figure 2). The computation cell width is incrementally increased and the corresponding free energy $\Delta$\textit{F}[ \textit{nk\textsubscript{b}T}] is calculated. The stretching energy penalty imposed on the BCP domains results in the semi-parabolic $\Delta$\textit{F}[ \textit{nk\textsubscript{b}T}] shape as a function of computation cell width with a minimum at the commensuration condition. Two $\Delta$\textit{F}[ \textit{nk\textsubscript{b}T}] curves are shown for two and three BCP domains, both resulting in L\textsubscript{0} = 3.71R\textsubscript{g}. The value of L\textsubscript{0} extracted from SCFT agrees well with the estimates from unit cell approximation UCA, where L\textsubscript{0} = 0.93L\textsubscript{B}\textsuperscript{32}. It is convenient to reference the geometry of the simulation to L\textsubscript{0} as it is the accessible measure in thin film experiments. 

\begin{figure}[h]
\centering\includegraphics[width=0.8\linewidth]{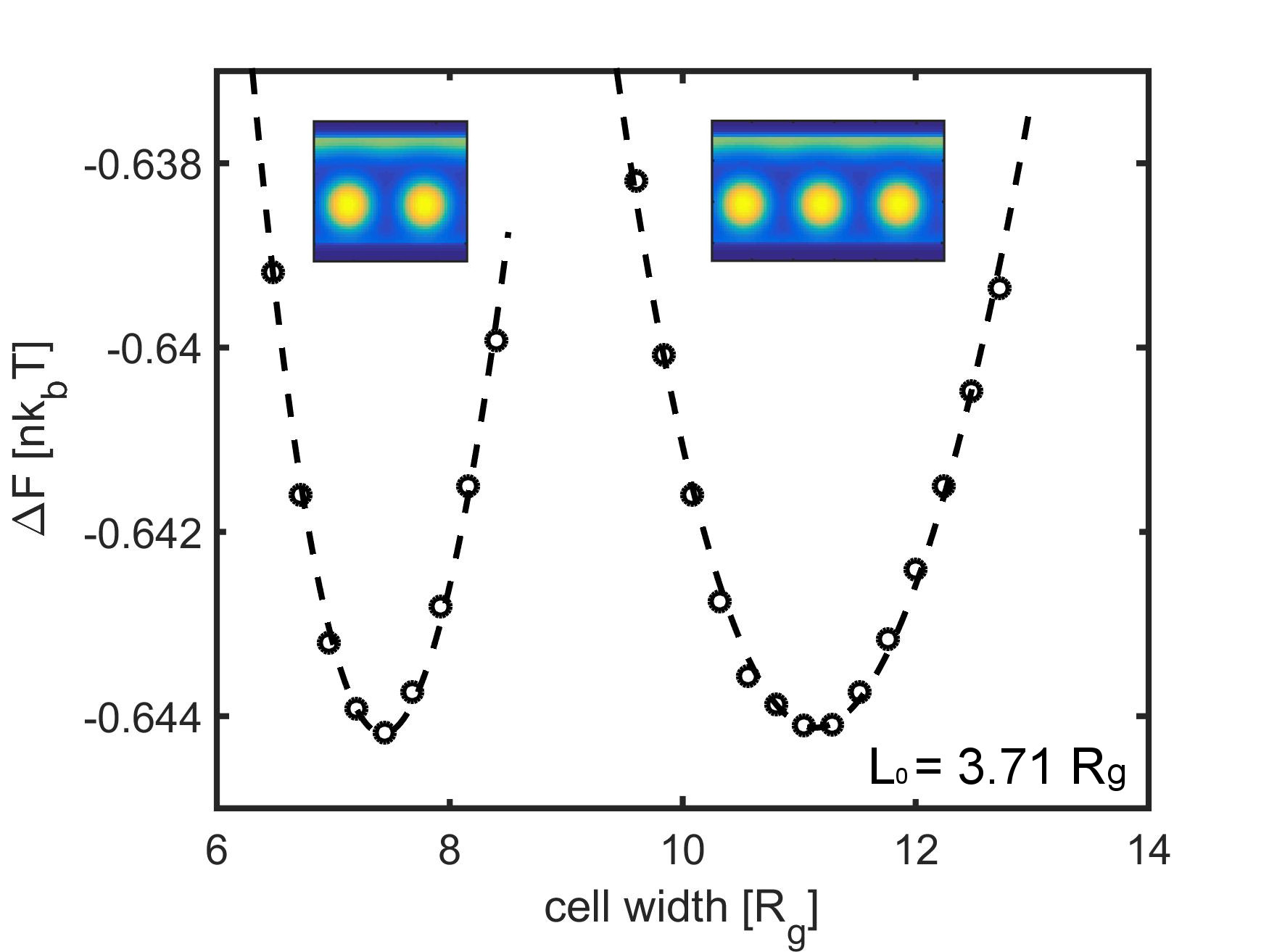}
\caption{Free energy plot for cylindrical BCP morphology as a function of cell width in a monolayer thin film. Equilibrium spacing L\textsubscript{0} = 3.71R\textsubscript{g} is smaller than bulk value L\textsubscript{B} = 3.96 R\textsubscript{g} due to the square symmetry imposed on the monolayer cylinders in thin film confinement.}
\end{figure}


\section{Topographic modulation of substrates }


Substrate modulation brings extra degrees of freedom that provide control over orientation and order of BCP domains. Modulation wavelength and surface chemistry would have a direct impact on the stretching and interfacial energies of BCP domains. In particular, chemical treatment of surfaces would increase their affinity to one of the polymer blocks creating wetting layers of one polymer near substrate surface. In addition, the magnitude of modulation spacing will apply a strain energy on BCP domains that is dependent on the degree of incommensurability between BCP periodicity and modulation spacing. Modulation shape adds extra complication to this argument. Extended modulation of substrate surfaces will amplify the effect of surface chemical treatment due to the excess surface area present. The crosstalk between modulation height and width will impose varying strain levels for different modulation geometries. 
An inverted parabolic barrier is used as a model system for the purpose of this study with the advantage of having a coupling between height and width that is controllable. The height of modulation can be arbitrarily set. A flat region between barriers is always present to resemble a substrate. The parabola has the expression \textit{y = h-s(x-x\textsubscript{0})\textsuperscript{2}} where \textit{h} is the barrier height and \textit{s} is related to the focal length of the parabola (\textit{x,y} only take positive values). Figure 3 shows a schematic of the computational cell projection. The BCP is confined along the vertical direction between a bottom substrate and a top free surface. The top free surface is set to have strong affinity to the minority block A that is fixed throughout all simulated cases (\textit{w\textsubscript{-}\textsuperscript{top}} = 4). All self-assembled domains lie in-plane. The chemical affinity of the substrate surface is a variable of the study to reveal the importance of surface functionalization. The width of the computational cell \textit{d} controls parabolic barrier separation given the periodic boundary conditions. Barrier separation would impose in-plane confinement strain energy at values incommensurate with L\textsubscript{0}. Finally, the parabolic barrier geometry is controlled by \textit{h} and \textit{s}. Two values of \textit{s} are used to control the spread of the parabola: 0.04 and 0.6 [1/pixels]. The height \textit{h} takes five values between 0.8R\textsubscript{g} and 2.4R\textsubscript{g} with a step of 0.4R\textsubscript{g}. Film thickness \textit{t} is kept at 1.5L\textsubscript{0} unless otherwise stated. Computational cell depth is fixed to L\textsubscript{0}. Parallel and perpendicular in-plane domains are seeded to the simulations and the corresponding free energies are calculated. It is important to note that strong top attraction suppresses the formation of vertically standing morphologies. This is true for a variety of BCP chemistries. However, our conclusions seem to be similar to those found with neutral boundary conditions\textsuperscript{25}.

\begin{figure}[h]
\centering\includegraphics[width=0.6\linewidth]{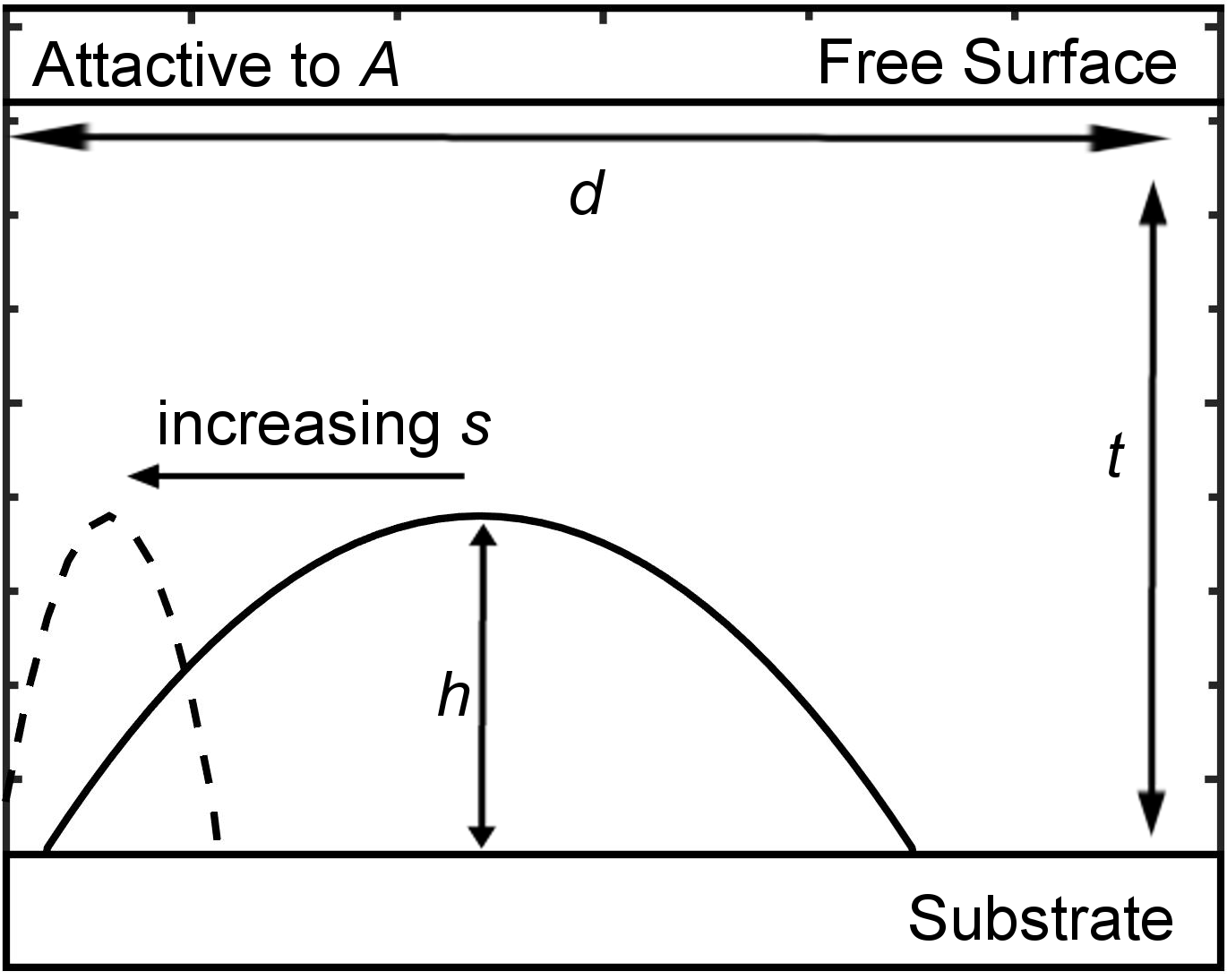}
\caption{Schematic of the computational cell for studying BCP cylindrical self-assembly on a modulated substrate. BCP cannot access top layer mimicking free surface, and bottom layer resembling a substrate. An inverted parabolic barrier is used that has variable spread \textit{s} and height \textit{h}. Top layer has a strong attraction to minority block \textit{A}, suppressing vertically standing structures (\textit{w\textsubscript{-}\textsuperscript{top}} = 4) . Bottom layer can have a weak preference to either of the blocks(\textit{w\textsubscript{-}} = $\pm$2) . The simulation is run for different cell width \textit{d}. Film thickness is \textit{t} = 1.5L\textsubscript{0} unless stated otherwise. }
\end{figure}

Figure 4 shows BCP domains aligned parallel to substrate modulation for different modulation height. A large barrier imposes additional strain on the BCP domains by limiting the accessible space to the polymer chains. More importantly, applying the same magnitude of surface affinity to either blocks (\textit{w\textsubscript{-}} = $\pm$2) shows a clear difference in self-assembly. A substrate that is attractive to the majority block B displaces domains A away from the barrier. At a large barrier height significant distortion of BCP domains is observed. A substrate attractive to minority block A exhibit a different behavior where there is a tendency of the BCP domains to merge with the barrier. At low barrier height, registry between polymer domains and barrier peaks is observed, resembling the behavior of chemoepitaxy. BCP domains shift to barrier side at a larger barrier height. 

\begin{figure}[h]
\centering\includegraphics[width=0.8\linewidth]{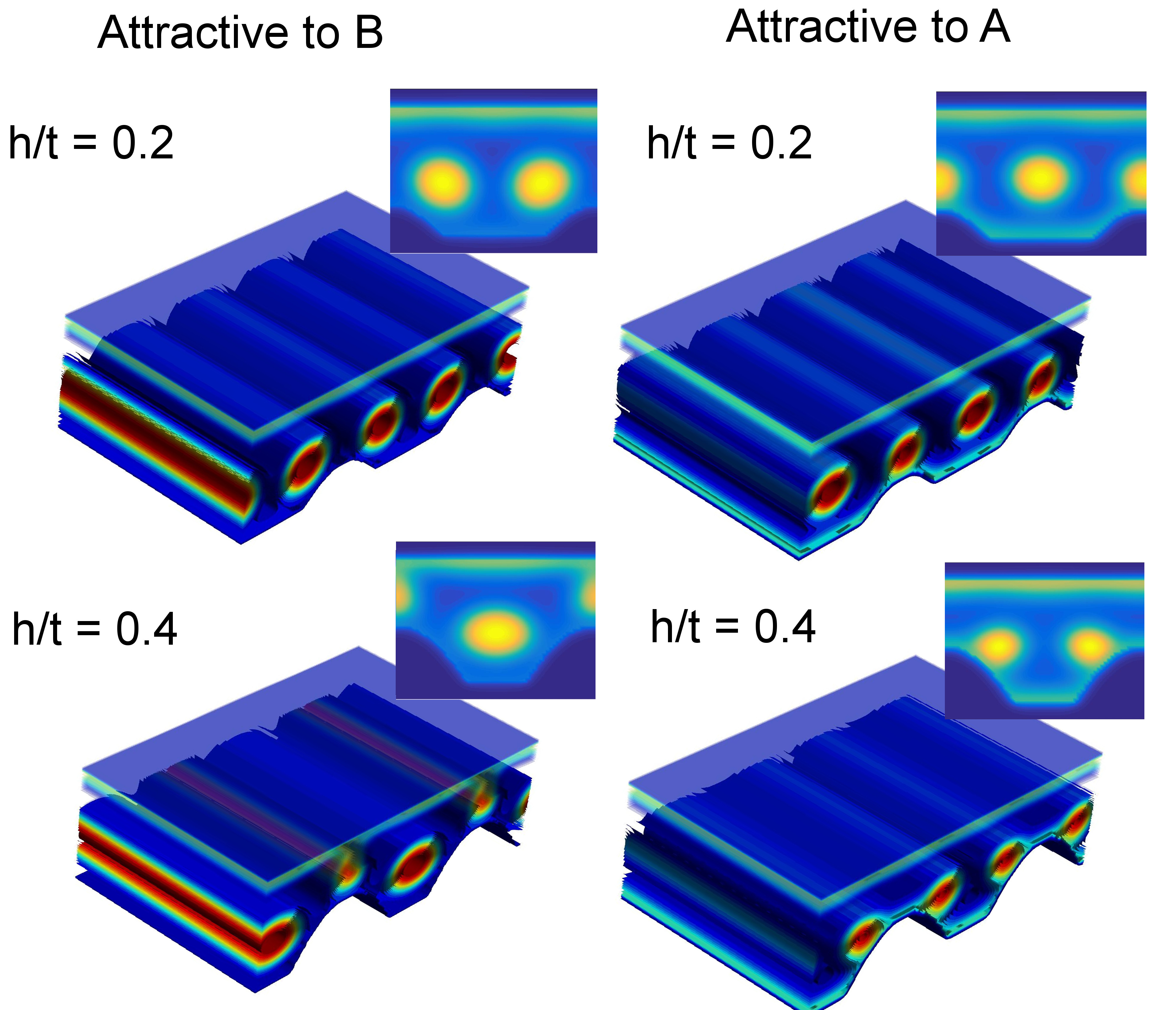}
\caption{3D Parallel morphology (inset shows 2D slice perpendicular to barrier) for low (\textit{h/t} = 0.2) and high (\textit{h/t} = 0.4) modulation for different levels of substrate affinity. Domains adsorb to topography when substrate is attractive to minority polymer \textit{A}, while domains are displaced by the polymer matrix when substrate is attractive to majority block \textit{B}.}
\end{figure}

The distinction in the behavior of self-assembled domains continues when aligned perpendicular to substrate modulation (see figure 5). Generally, cylinders crossing over barriers tend to curve upwards away from the substrate in \textit{B} preferential substrates. This is understood by the fact that block \textit{B} conforms to the substrate, displacing block \textit{A}. On the other hand, BCP domains tend to fuse with the barrier for A preferential substrates. This introduces a small curvature towards the substrate in the vicinity of the barrier. Interestingly, the interaction of BCP domains with substrate modulation determines the ability to produce continuous bridges across barriers. In figure 6, a plot of BCP density \textit{A} as a function of film thickness located at barrier peak is demonstrated. In case of a \textit{B} preferential substrate, a peak in density \textit{A} is observed till barrier height \textit{h/t} = 0.3. A higher barrier causes the peak to disappear indicating a disconnection between BCP domains on both sides of the barrier. Peak density vanishes at a lower barrier height (\textit{h/t} = 0.2) for an \textit{A} affine substrate. The ability of the BCP domains to merge with the barrier minimizes the energy penalty of creating an interface at domain termination, favoring domain disconnection at lower barrier height. These observations demonstrate the role substrate chemical treatment has on the resulting morphology of BCP domains. 

\begin{figure}[h]
\centering\includegraphics[width=0.8\linewidth]{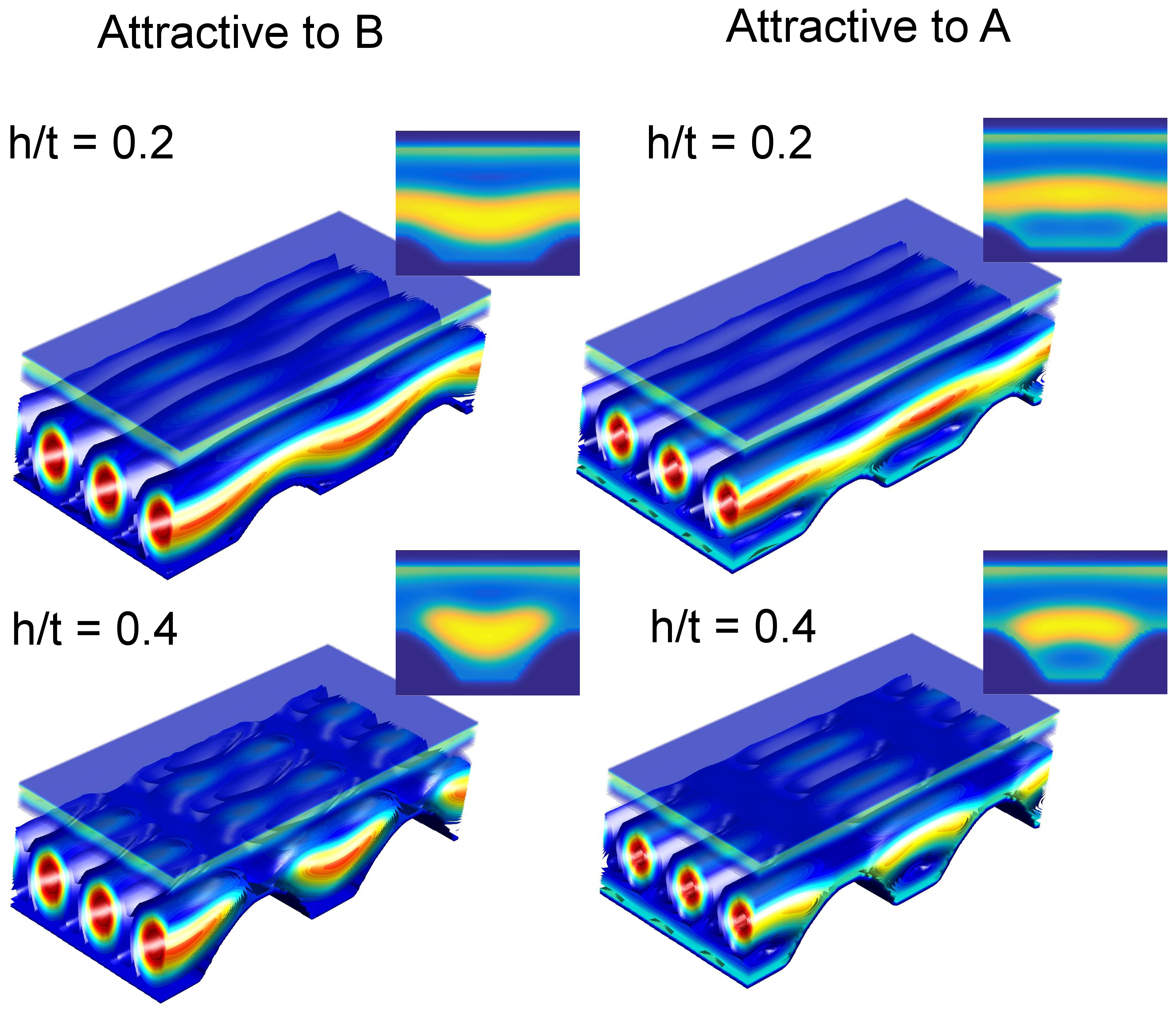}
\caption{3D perpendicular morphology (inset shows 2D slice) for low (\textit{h/t} = 0.2) and high (\textit{h/t} = 0.4) modulation for different substrate affinity. BCP domains tend to curve towards substrate and fuse with substrate undulation in minority \textit{A} affine case. When substrate is attractive to majority polymer \textit{B}, domains tend to curve away from topography.}
\end{figure}

\begin{figure}[h]
\centering\includegraphics[width=0.8\linewidth]{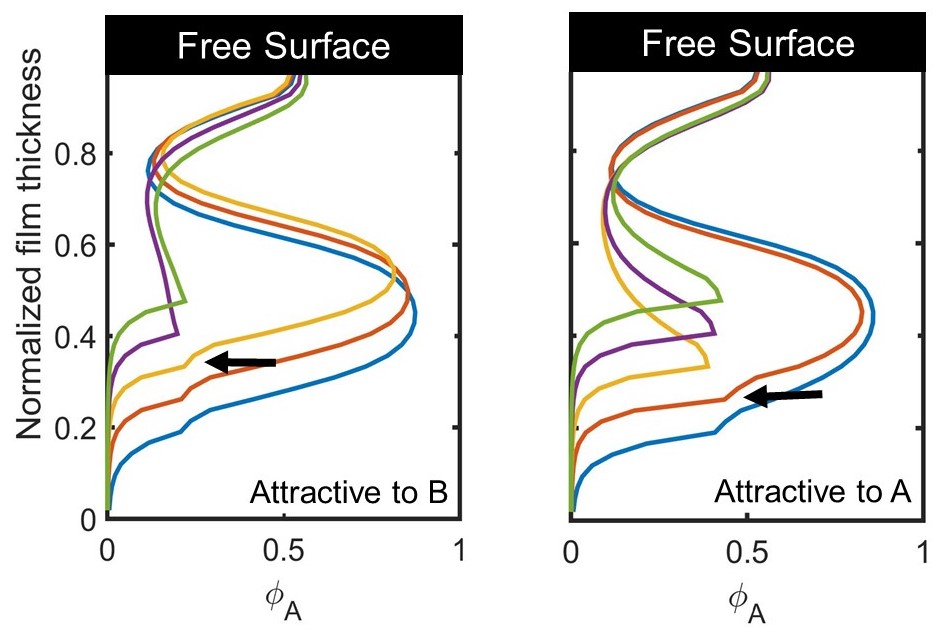}
\caption{ Minority block \textit{A} density sliced at the peak of substrate modulation. Bridging across the substrate barrier persists to large \textit{h} when substrate is attractive to majority block. For minority affine substrates, domains get disconnected across barriers as they fuse with the wetting layer on the substrate, which minimizes the energy penalty of domain termination. }
\end{figure}

As was discussed, substrate modulation imposes additional stretching energy on BCP chains due to coupling between height and width of barrier. Figure 7 shows phase diagrams of parallel and perpendicular orientations for varying normalized barrier spacing and normalized barrier height. Light areas are stable for parallel orientations. For thin barriers (\textit{s} = 0.6), weak coupling between height and width of barrier is expected. Parallel orientation is stabilized in the vicinity of commensuration. This is shown through the periodic striped domains in the phases diagram at L = \textit{m}L\textsubscript{0} where \textit{m} is an integer. In addition, the effect of confinement diminishes for wider barrier spacing as reflected by the continuous decay in $\Delta$\textit{F}[\textit{nk\textsubscript{b}T}] at large \textit{m}. Parallel and perpendicular orientation is expected to be degenerate at large values of \textit{m}. While commensuration signature is noted in the phase diagram, a much more pronounced effect of substrate affinity is observed. Stable parallel domains, effectively independent of barrier height, are attained at commensuration for a \textit{B} preferential substrate. The perpendicular orientation prevails in the case of \textit{A} preferential substrates independent of barrier spacing for barrier heights \textit{h/t} $<$ 0.3. Shallow stable parallel regions emerge at 0.3$<$\textit{h/t}$<$0.5 near commensuration. Similar phase behavior is obtained at a weaker BCP substrate interaction (\textit{w\textsubscript{-}} = $\pm$1), see figure (S1). Increasing barrier width (\textit{s} = 0.04) further amplifies the effect of substrate affinity. Discontinuous regions of stable parallel domains emerge depending on \textit{h}. Weak stability of parallel orientation is observed at \textit{h/t} $\sim$ 0.2 at barrier spacing \textit{d} close to commensurability. Strong stable parallel orientation (large free energy difference) re-emerges for high barrier 0.3$<$\textit{h/t}$<$0.5, but displaced from integer multiple of L\textsubscript{0} due to the physical size of barrier. On the other hand, perpendicular orientation completely dominated the phase diagram for the range of barrier height and spacing explored in the study for an \textit{A} preferential substrate. This behavior was also observed at film thickness of 1.7L\textsubscript{0} that still maintains a monolayer of cylinders, see figure S2.
The results of SCFT show the non-trivial interaction between stretching energy imposed by barrier confinement and the interfacial component of energy due to substrate affinity. It is evident that there is a strong asymmetry in the effect of interfacial energy, where numerically applying the same magnitude of attraction to either blocks causes the polymer domains to respond differently. This can be rationalized by the asymmetry in BCP composition. Having the substrate affine to the majority polymer \textit{B} can be accommodated by the continuous polymer matrix, while affinity to the minority polymer \textit{A} will have an energy penalty due to the presence of both blocks in the vicinity of a surface, a behavior that has been reported elsewhere\textsuperscript{33}. In addition, SCFT signifies the role of extended barriers. Separating wide barriers at a spacing commensurate with BCP natural periodicity still applies a stretching energy penalty throughout film thickness. This is strongly dependent on barrier shape and height.  With shallow barriers, distributing integer number of polymer domains on an uneven surface proves unfavorable and polymer domains prefer to align normal to modulation direction where there is no constraint on BCP domain separation. 

\begin{figure}[h]
\centering\includegraphics[width=0.7\linewidth]{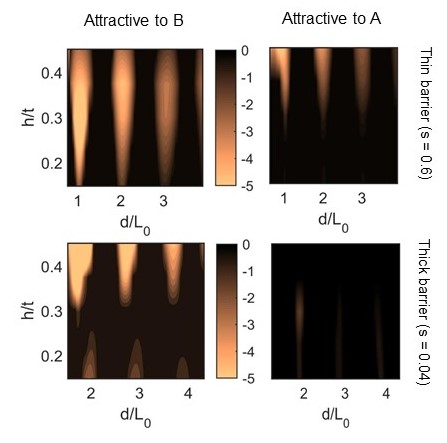}
\caption{Phase diagram of regions of stable parallel orientation calculated by \textit{F\textsubscript{$\parallel$}-F\textsubscript{$\perp$}} [\textit{10\textsuperscript{3}nk\textsubscript{b}T}] for different cell width and barrier height. Light areas are stable for parallel orientations. The majority attractive substrate shows stable parallel regions near commensurate condition especially for thin barriers. Minority attractive substrate shows large window of stability for perpendicular orientation unless high barriers are present. Still, wide barriers cause the perpendicular orientation to dominate the phase diagram independent of barrier spacing.(\textit{t}= 1.5L\textsubscript{0})  }
\end{figure}


\section{Analytical model using SST}

Substrate modulation applies non-trivial boundary confinement over BCP thin-film. Nonetheless, invoking a round unit cell approximation (UCA), one can reasonably capture the essence of the problem and give insights to the major factors governing BCP self-assembly. In this work, UCA is attained by assuming that the exact repeating geometric cell surrounding a single polymer cylindrical domain can be approximated by a cylindrical cell with an equal volume to that accessible to the polymer. By doing so, the strong segregation theory SST analysis is greatly simplified. As film thickness is kept constant during the analysis, the round unit cell has a constant radius along film thickness, while radius parallel to substrate changes depending on barrier spacing. Hence, the round unit-cell deforms into an ellipse when stretched/compressed for cylinders aligning parallel to substrate modulation. 
Here, an incompressible \textit{AB} diblock melt composed of \textit{n} chains. The polymer has a density  \textit{$\rho$\textsubscript{0} = nN/V} where  \textit{V} is the volume occupied by the polymer. Each monomer occupies a volume  \textit{v} and has a statistical segment length  \textit{a}. The chain has a radius of gyration R\textsubscript{g} = $\sqrt{(Na^2/6)}$. 
In the limit of strong segregation, it is assumed that the core consists of almost exclusively block \textit{A}, while the surrounding corona is block \textit{B}. A narrow interface is formed where the copolymer junction is confined. Chains are extended so the effect of their fluctuation about their mean path is assumed to be negligible. Each polymer domain occupies a cylindrical volume with outer radius \textit{R} and radius of the cylindrical interface \textit{R\textsubscript{A}} = $\sqrt{f}$ \textit{R}. This condition ensures that the inner domain  \textit{A} occupies  \textit{f} portion of the total volume of the cylindrical cell. In SST, L\textsubscript{B} can be estimated as\textsuperscript{32, 34} 

\begin{equation}
\label{eq:emc}
L_B = \sqrt{\frac{\pi}{\frac{1}{2}\sqrt{3}}}R_0;\;\;\;\; 
R_0^3 = \frac{2\sqrt{f}\sqrt{\chi N}}{\frac{\pi^2}{48}+\frac{\pi^2}{4}
( f/2-2\sqrt{f}/3-f^2/12+1/4)}R_g^3
\end{equation}

The reference geometry of the BCP domains are obtained from a monolayer of in-plane cylinders on a flat substrate. Every domain occupies a square unit cell of side length \textit{d\textsubscript{0}} = $\sqrt{\pi}R_0$. The presence of a parabolic barrier reduces the physical space available to the polymer domain. The parabolic profile is expressed as follows

\begin{equation}
\label{eq:emc}
y =
  \begin{cases}
    h(1-(\frac{x}{x_0}-1)^2)       & \quad 0<x<2x_0\\
    0  & \quad 2x_0<x<d
  \end{cases}
\end{equation}

 The accessible cross-sectional area \textit{A\textsubscript{c}} for a polymer domain parallel to a barrier is

\begin{equation}
\label{eq:emc}
A_c[d] = dd_0-\frac{4}{3}hx_0
\end{equation}

Where \textit{d} is the barrier spacing. The BCP domain is assumed to deform into an elliptical shape when stretched/compressed, with a radius normal to substrate \textit{R\textsubscript{n}}\textsuperscript{$\parallel$} $\equiv$ \textit{R\textsubscript{0}}, and parallel to substrate \textit{R\textsubscript{p}}\textsuperscript{$\parallel$} =  $A_c[d]/\pi R_n^\parallel$

The same logic follows when BCP domains are normal to barrier; however, the BCP domain is stress free in the direction parallel to barrier, and gets deformed into an ellipse along film thickness in the vicinity of the barrier, \textit{R\textsubscript{p}}\textsuperscript{$\perp$} $\equiv$ \textit{R\textsubscript{0}} and $R_n^\perp (x) = d_0(d_0-y(x))/(\pi R_p^\perp)$. Hence, a BCP domain has a circular cross-section in the region of flat substrate, and only deforms when crosses over the barrier (see figure 8). 

\begin{figure}[h]
\centering\includegraphics[width=0.8\linewidth]{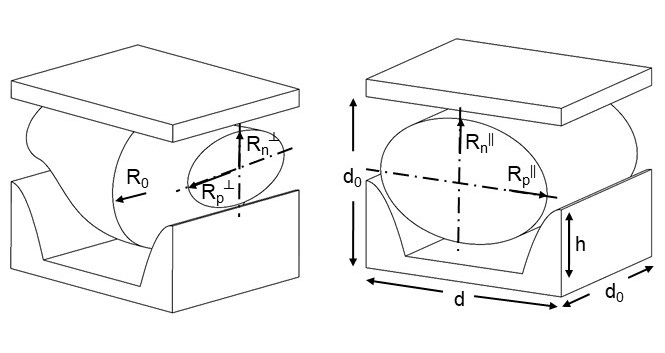}
\caption{Schematic showing parameters employed in calculating free energies of parallel and perpendicular orientations in the limit of SST.}
\end{figure}

Having set the geometric characteristics of the system, stable phases can be determined using free energy calculations. The free energy expression \textit{F} contains the elastic chain stretching components of the core and corona chains: $F_A$ and $F_B$, respectively. An interfacial free energy $F_I$ is calculated for the core-corona interface. In addition, surface effects need to be considered in case of thin films. We only focus on the substrate surface as the top free surface is common in both perpendicular and parallel domain orientations and drops out when comparing \textit{F}. Hence

\begin{equation*}
    F^\parallel = F_A^\parallel+F_B^\parallel+F_I^\parallel+\gamma \Sigma 
\end{equation*}

\begin{equation*}
    \frac{F_I^\parallel}{nk_bT} = 2 \sqrt{f} \sqrt{\chi N}
    \frac{\sqrt{(R_p^{\parallel 2}+R_n^{\parallel 2})/2}}{R_p^{\parallel}R_n^{\parallel}}R_g
\end{equation*}

\begin{equation*}
    \frac{F_A^\parallel}{nk_bT} =\frac{3\pi^2}{48f^2R_g^2V} \int_A dzA(z)z^2
\end{equation*}

\begin{equation}
    \frac{F_B^\parallel}{nk_bT} =\frac{3\pi^2}{48(1-f)^2R_g^2V} \int_B dzA(z)z^2
\end{equation}

The stretching energy is calculated by discretizing the core and corona into concentric ellipses and integrating radially (along radial coordinate \textit{z}) the circumferential area of each ellipse. The presence of barrier is accounted for using the additional surface area $\Sigma$ of a parabola, and interfacial tension $\gamma [nk_bT] = R_g/V\sqrt{\chi _sN}$ which penalizes interaction with block \textit{B}.

Similarly, free energy of perpendicular domains is calculated taking into account the change of BCP cross sectional shape due to the presence of a barrier. 

\begin{equation*}
    F^\perp = F_A^\perp+F_B^\perp+F_I^\perp+\gamma \Sigma (1-2f) 
\end{equation*}

\begin{equation*}
    \frac{F_I^\perp}{nk_bT} = \sqrt{f} \sqrt{\chi N}
    \frac{C}{V}R_g
\end{equation*}

\begin{equation*}
    \frac{F_A^\perp}{nk_bT} =\frac{3\pi^2}{48f^2R_g^2V} \int_0^d \int_A dxdzA(x,z)z^2
\end{equation*}

\begin{equation}
    \frac{F_B^\perp}{nk_bT} =\frac{3\pi^2}{48(1-f)^2R_g^2V} \int_0^d \int_B dxdzA(x,z)z^2
\end{equation}

Where \textit{x} is a coordinate along the direction of barrier spacing, \textit{C} is the interfacial area between the two blocks (numerically calculated). Note that the effect of $\Sigma$ is reduced by $(1-2f )$ due to the interaction of the minority block \textit{A} with an affine substrate. It is assumed that the substrate has equal interfacial tension with each block $\gamma$ = $\gamma _B$ = -$\gamma _A$. Block dependent interaction can be employed; however, it would not change the underlying physics of the problem.

Figure 9a shows the variation of $F[nk_bT]$ as a function of barrier spacing \textit{d} for a system with no substrate interaction ($\chi _sN$ = 0). BCP domains that are parallel to barrier show the semi-parabolic plot of $F^\parallel$ with a minimum in free energy where interfacial and stretching energies balance. In fact, it is seen that changing cell width \textit{d} effectively increases domain size. Growing system volume reduces the density of surfaces separating both blocks causing a reduction in interfacial energy per volume ($F_I \sim 1/R$). Hence, interfacial energy promotes large domain size and spacing. On the contrary, stretching energy increases with growing system size as BCP chains have to extend radially outwards to fill the expanding space. The extended chain conformation reduces chain entropy and increases stretching energy ($F_A+ F_B \sim R^2$). These two competing effects balance near $d_0 = L_0$. On the other hand, BCP domains perpendicular to barrier experience negligible reduction in $F^\perp$ as domain size is not affected by barrier spacing, except near barrier. The free energy difference $F^\parallel - F^\perp$ can then be employed to reveals the regions of stability of different morphologies. Figure 9b presents a phase diagram of parallel/perpendicular morphology as a function of barrier spacing \textit{d} and barrier height \textit{h} for a wide barrier ($x_0 = d_0/5$). Similar to SCFT results, stability of parallel domains emerges with the increase of barrier height. In addition, equilibrium barrier spacing drifts towards larger values of \textit{d} due to the strong coupling between barrier height and width. The decrease of accessible space for BCP domains at large barrier heights causes the minimum of $F^\parallel - F^\perp$ to shift towards large \textit{d}. The phase diagram reveals the spread of stable regions of parallel orientation at large \textit{h} and wide barriers. Furthermore, substrate affinity has a strong effect in limiting the region of stability of parallel morphology. Figure 9c shows partitioning contour bounding region of stable parallel domains (arbitrarily fixed at $F^\parallel - F^\perp = -10^{-3}[nk_bT]$ for different values of $\chi _sN$ (marked on each contour plot in figure 9c). It is observed that substrate affinity can significantly extend the region of stability of perpendicular orientation and confine it to large barrier height. Note that the numerical values of $\chi _sN$ at which orientation switches from parallel to perpendicular depends on how substrate interaction is accounted for; however, this does not change the overall trend observed in the free energy plots. 

\begin{figure}[h]
\centering\includegraphics[width=0.8\linewidth]{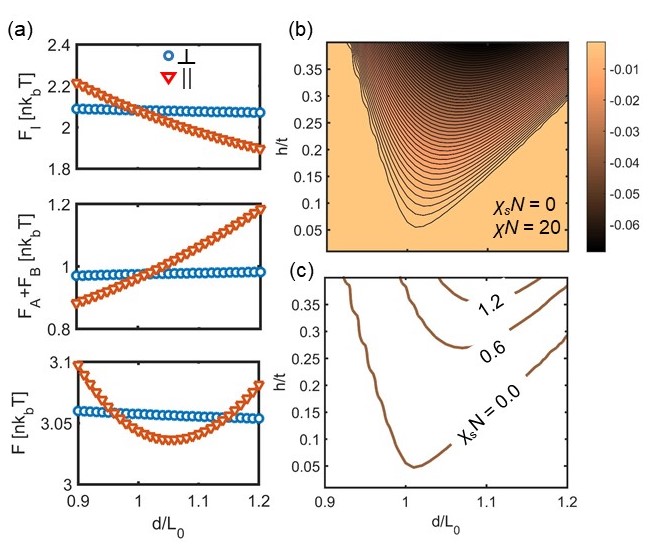}
\caption{(a) Sample free energy calculations of parallel and perpendicular orientations including interfacial $F_I$ and stretching energies $F_A+F_B$. Increasing cell width reduces $F_I$, but increases  $F_A+F_B$ for parallel orientation. The perpendicular orientation shows slight reduction in $F$ with increasing cell width as the BCP domain shape is not altered except near barrier. (b) Phase diagram of stable parallel orientation calculated by $F^\parallel - F^\perp$. (c) Parallel orientation border at ($-10^{-3}[nk_bT]$) for different levels of substrate repulsion to majority block \textit{B}. The higher the repulsion, the large the stable region of perpendicular orientation.}
\end{figure}

\section{Results and discussion}

The findings of SCFT and SST analytical model provide a comprehensive view of the conditions producing in-plane parallel or perpendicular monolayer cylindrical structures. The general guidelines that can be drawn from the analysis show that parallel orientation is accessible in majority preferential substrates near regions of commensuration between barrier separation and polymer equilibrium spacing. This is particularly relevant for thin barrier shapes. Higher barrier heights with respect to monolayer thin-films are more important to achieve parallel orientation for extended barriers. Perpendicular orientation is accessible for substrates that are weakly attractive to minority block with shallow substrate modulation. Extended barriers increase the range of stability of perpendicular orientation, with limited effect of barrier spacing. The current work additionally reveals the coupling between barrier height and width. Despite the limited experimental work on systems similar to what is studied in this work, analogies can still be drawn from the literature. In-plane cylinders forming on a saw-tooth modulated substrate exhibited film-thickness dependent orientation \textsuperscript{15, 18}. In-plane cylinders perpendicular to ridges were formed at large film thicknesses on both sapphire and silicon saw-toothed substrates (\textit{t}$\sim$ 1.5L\textsubscript{0}). When film thickness was decreased (\textit{t}$\sim$ L\textsubscript{0}) parallel orientation was obtained\textsuperscript{18}. Similar, behavior was reported on shallow trench patterned substrates\textsuperscript{35}. Full orthogonal alignment was obtained at thicker films $>$ 1.7L\textsubscript{0}. Capillary flow into trenches during solvent evaporation was proposed as a potential cause for obtaining perpendicular orientation\textsuperscript{35}. However, BCP self-assembled on very shallow trenches (15nm deep) makes capillary effect highly implausible, especially at thicker films. In addition, orientation should evolve upon annealing if it was a metastable state driven by capillary forces\textsuperscript{36}. This is yet to be experimentally demonstrated. We argue that the perpendicular orientation obtained on shallow trenches, as well as saw-tooth substrates at thicker films emerged given the untreated substrates (that can show weak preferentiality to either blocks) and shallow barrier height due to the increased film thickness. More importantly, our phase diagram explains well, previous experimental results in \textsuperscript{23} where multi-layer cylinders aligned into a mesh structure. The untreated, lower extended barrier (cylindrical topography) managed to template the top monolayer of low molecular weight cylinders into perpendicular orientation. 

Recent experimental work on a similar system demonstrated the ability of the top layer to align parallel to the bottom one\textsuperscript{24}. It was argued that orientation manipulation is achieved through control over film thickness. While no exact film thickness measurements were conducted, change in brightness in the final scanning electron microscopy SEM micrographs were employed to qualitatively indicate variations in film thickness. The accompanied analysis using SCFT concluded that incommensurability between the two layers dictated orientation preference. There, it was argued that orientation normal to the bottom layer was stabilized when the hexagonal packing of parallel orientation could not be accommodated within the film. However, SCFT results had little resemblance to the accompanied experimental work. For example, the perpendicular orientation obtained by SCFT for spacing ratio of 1.5 (Fig. 2c\textsuperscript{24}) depends on forming segmented BCP domains in between the barriers that are further used to align the top cylindrical layer. This is missing when examining the experimental SEM scan in Fig. 4a\textsuperscript{24}. Further, the SCFT morphology for spacing ratio of 2 (Fig. 3b\textsuperscript{24}), which is used to justify the importance of commensurability, shows the cylindrical top layer in perfect registration with the half cylindrical barrier on substrate. Again, this is not observed in the SEM scans in Fig. 4b\textsuperscript{24}. In fact, examining Fig. 4b\textsuperscript{24} reveals the fact that domains form thick bands of triple cylinders where the top two cylinders adsorb on the lower topographic one. This triple cylinder thick band is nicely captured in our SCFT work as was shown in figure 4 \textit{(h/t = 0.4, minority \textit{A} attractive substrate)}. Finally, it was stated that switching orientation between parallel and perpendicular was due to the increase in top film thickness. Evidence in the article was provided through the variation in brightness between different regions of SEM scans. It can be argued that dark regions might be a result of a thicker bottom layer rather than top layer, due to uneven plasma etching or boundary effects during spin coating. It would be instructive to provide quantitative local film thickness measurements to further elucidate the role of film thickness in the aforementioned experiments.

\section{Conclusion}

In summary, we have explored the self-assembly of cylinder forming BCPs on modulated substrates. Our results explored the large parameter space governing the final orientation and order of the self-assembled monolayer using both SCFT and SST. We showed that perpendicular orientation of top layer can be achieved using shallow substrate topography with weak substrate attraction to the minority block. Barrier spacing seems to play a minor role indicating the limited effect of commensurability between barrier spacing and BCP equilibrium periodicity. On the other hand, large substrate corrugations in the form of large barrier height promoted parallel orientation in the vicinity of commensurations. This was particularly true when substrate is attractive to majority polymer. 

The analysis revealed the asymmetric effect of substrate functionalization on the final structure. Such behavior was rationalized due to the asymmetry of block concentration. Weak substrate affinity was the focus of this work guided by the body of experimental work that typically employ non-treated substrates for subsequent layer deposition. In addition to the conditions explored in this work, other factors might add to the rich behavior of BCP self-assembly such as film thickness, and magnitude and distribution of substrate affinity.

Our work should be important for better understanding the alignment of block copolymer domains in topographic features as well as in the design of newer 3D films that build on previous block copolymer layers.

\section{References}
\setlength{\parindent}{0ex}

1.	Darling, S. Progress in Polymer Science 2007, 32, (10), 1152-1204.

2.	Matsen, M. W. The European Physical Journal E 2009, 30, (4), 361.

3.	Bita, I.; Yang, J. K.; Jung, Y. S.; Ross, C. A.; Thomas, E. L.; Berggren, K. K. Science 2008, 321, (5891), 939-943.

4.	Cheng, J. Y.; Mayes, A. M.; Ross, C. A. Nature materials 2004, 3, (11), 823.

5.	KG, A. T.; Gotrik, K.; Hannon, A.; Alexander-Katz, A.; Ross, C.; Berggren, K. Science 2012, 336, (6086), 1294-1298.

6.	Tavakkoli KG, A.; Hannon, A. F.; Gotrik, K. W.; Alexander‐Katz, A.; Ross, C. A.; Berggren, K. K. Advanced Materials 2012, 24, (31), 4343-4343.

7.	Gadelrab, K. R.; Ding, Y.; Pablo-Pedro, R.; Chen, H.; Gotrik, K. W.; Tempel, D. G.; Ross, C. A.; Alexander-Katz, A. Nano letters 2018.

8.	Kim, S. O.; Solak, H. H.; Stoykovich, M. P.; Ferrier, N. J.; de Pablo, J. J.; Nealey, P. F. Nature 2003, 424, (6947), 411.

9.	Stoykovich, M. P.; Kang, H.; Daoulas, K. C.; Liu, G.; Liu, C.-C.; de Pablo, J. J.; Müller, M.; Nealey, P. F. Acs Nano 2007, 1, (3), 168-175.

10.	Stoykovich, M. P.; Müller, M.; Kim, S. O.; Solak, H. H.; Edwards, E. W.; De Pablo, J. J.; Nealey, P. F. Science 2005, 308, (5727), 1442-1446.

11.	Angelescu, D. E.; Waller, J. H.; Adamson, D. H.; Deshpande, P.; Chou, S. Y.; Register, R. A.; Chaikin, P. M. Advanced Materials 2004, 16, (19), 1736-1740.

12.	Chen, Z.-R.; Kornfield, J. A.; Smith, S. D.; Grothaus, J. T.; Satkowski, M. M. Science 1997, 277, (5330), 1248-1253.

13.	Nicaise, S. M.; Gadelrab, K. R.; KG, A. T.; Ross, C. A.; Alexander-Katz, A.; Berggren, K. K. Nano Futures 2018, 1, (3), 035006.

14.	Majewski, P. W.; Yager, K. G. ACS nano 2015, 9, (4), 3896-3906.

15.	Hong, S. W.; Huh, J.; Gu, X.; Lee, D. H.; Jo, W. H.; Park, S.; Xu, T.; Russell, T. P. Proceedings of the National Academy of Sciences 2012, 109, (5), 1402-1406.

16.	Park, S.; Lee, D. H.; Xu, J.; Kim, B.; Hong, S. W.; Jeong, U.; Xu, T.; Russell, T. P. Science 2009, 323, (5917), 1030-1033.

17.	Lee, D.-E.; Ryu, J.; Hong, D.; Park, S.; Lee, D. H.; Russell, T. P. ACS nano 2018, 12, (2), 1642-1649.

18.	Hong, S. W.; Voronov, D. L.; Lee, D. H.; Hexemer, A.; Padmore, H. A.; Xu, T.; Russell, T. P. Advanced Materials 2012, 24, (31), 4278-4283.

19.	Aissou, K.; Shaver, J.; Fleury, G.; Pécastaings, G.; Brochon, C.; Navarro, C.; Grauby, S.; Rampnoux, J. M.; Dilhaire, S.; Hadziioannou, G. Advanced Materials 2013, 25, (2), 213-217.

20.	Rahman, A.; Majewski, P. W.; Doerk, G.; Black, C. T.; Yager, K. G. Nature communications 2016, 7, 13988.

21.	Ruiz, R.; Sandstrom, R. L.; Black, C. T. Advanced Materials 2007, 19, (4), 587-591.

22.	Majewski, P. W.; Rahman, A.; Black, C. T.; Yager, K. G. Nature communications 2015, 6, 7448.

23.	KG, A. T.; Nicaise, S. M.; Gadelrab, K. R.; Alexander-Katz, A.; Ross, C. A.; Berggren, K. K. Nature communications 2016, 7, 10518.

24.	Carpenter, C. L.; Nicaise, S.; Theofanis, P. L.; Shykind, D.; Berggren, K. K.; Delaney, K. T.; Fredrickson, G. H. Macromolecules 2017, 50, (20), 8258-8266.

25.	Zhu, Y.; Aissou, K.; Andelman, D.; Man, X. arXiv preprint arXiv: 1811.01539 2018.

26.	Semenov, A. Zh. Eksp. Teor. Fiz 1985, 88, (4), 1242-1256.

27.	Matsen, M. The Journal of chemical physics 1997, 106, (18), 7781-7791.

28.	Geisinger, T.; Müller, M.; Binder, K. The Journal of chemical physics 1999, 111, (11), 5241-5250.

29.	Tsori, Y.; Andelman, D. The European Physical Journal E 2001, 5, (1), 605-614.

30.	Abate, A. A.; Vu, G. T.; Pezzutti, A. D.; García, N. A.; Davis, R. L.; Schmid, F.; Register, R. A.; Vega, D. A. Macromolecules 2016, 49, (19), 7588-7596.

31.	Knoll, A.; Tsarkova, L.; Krausch, G. Nano letters 2007, 7, (3), 843-846.

32.	Helfand, E.; Wasserman, Z. Macromolecules 1980, 13, (4), 994-998.

33.	Bosse, A. W.; Garcia-Cervera, C. J.; Fredrickson, G. H. Macromolecules 2007, 40, (26), 9570-9581.

34.	Grason, G. M. Physics reports 2006, 433, (1), 1-64.

35.	Choi, J.; Li, Y.; Kim, P. Y.; Liu, F.; Kim, H.; Yu, D. M.; Huh, J.; Carter, K. R.; Russell, T. P. ACS applied materials and interfaces 2018, 10, (9), 8324-8332.

36.	Cheng, L.-C.; Gadelrab, K. R.; Kawamoto, K.; Yager, K. G.; Johnson, J. A.; Alexander-Katz, A.; Ross, C. A. Nano letters 2018.

\newpage
\renewcommand\thefigure{S\arabic{figure}}    
\setcounter{figure}{0}    

\section{Supplementary }

\begin{figure}[h]
\centering\includegraphics[width=0.4\linewidth]{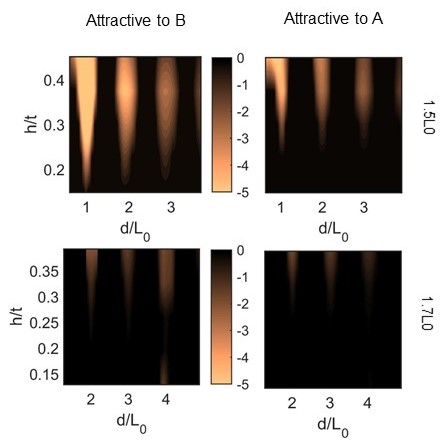}
\caption{ Phase diagram of regions of stable parallel orientation calculated by \textit{F\textsubscript{$\parallel$}-F\textsubscript{$\perp$}} [\textit{10\textsuperscript{3}nk\textsubscript{b}T}] for different cell width and barrier height ($w_-$ = 1 and \textit{s} =0.6)}
\end{figure}

\begin{figure}[ht]
\centering\includegraphics[width=0.4\linewidth]{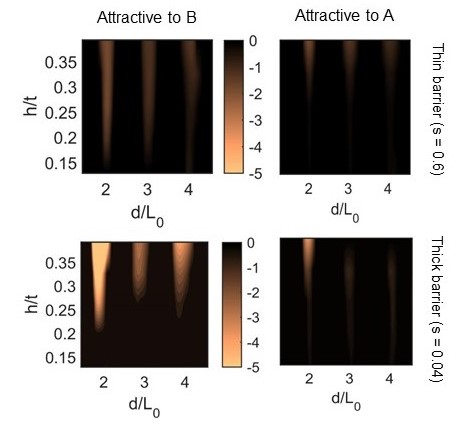}
\caption{ Phase diagram of regions of stable parallel orientation calculated by \textit{F\textsubscript{$\parallel$}-F\textsubscript{$\perp$}} [\textit{10\textsuperscript{3}nk\textsubscript{b}T}] for different cell width and barrier height (\textit{t} = 1.7L$_0$, $w_-$ = 2)}
\end{figure}











\end{document}